\documentstyle[12pt]{article}
\begin{document}
{\hskip 12.cm} SNUTP 96-121\par
\vspace{1ex}
\vspace{5ex}
\begin{center}        
{\LARGE \bf Some Remarks on the Baryon-Meson Couplings in the $1/N_c$
Expansion}\\
\vspace{7ex}
{\sc Chun Liu}\footnote{\it email: liuc@ctp.snu.ac.kr}\\
\vspace{3ex}     
{\it Center For Theoretical Physics, Seoul National University}\\
{\it Seoul, 151-742, Korea}\\

\vspace{10.0ex}
{\large \bf Abstract}\\
\vspace{4ex}
\begin{minipage}{130mm}
                                                                               
   The original results for the baryon-pion couplings in the large $N_c$
QCD can be understood in a simpler way in the Hartree-Fock picture.  
The large $N_c$ relation and its $1/N_c$ correction between the heavy 
baryon-meson coupling and the light baryon-meson coupling are emphasized.
Application to the baryon-$\rho$ meson interactions is 
straightforward.  The implications of recent experimental result for the 
strong coupling constants of the heavy baryon chiral lagrangian are 
discussed.
\par
\vspace{2.0cm}
\end{minipage}
\end{center}

\newpage
                                                                               
   Baryons provide us testing ground to Quantum Chromodynamics (QCD).
Experiments have collected rich data for light baryons, and are 
accumulating more and more data for heavy baryons.  For the theoretical 
calculation, the task is to apply model-independent methods of the 
non-perturbative QCD.  The tools which fit the task are like lattice QCD,
$1/N_c$ expansion, chiral lagrangian, heavy quark effective theory (HQET), 
QCD sum rules and so on.  In this paper, we discuss some aspects in the 
$1/N_c$ expansion of baryon-meson strong coupling constants.  
In the large $N_c$ limit [1], the meson properties 
can be obtained by analyzing planar diagram, and baryon properties by
considering Hartree-Fock picture.  Hadrons can be understood qualitatively
quite well.  
\par
\vspace{1.0cm}    
   Recently there are renewed interests in the $1/N_c$ expansion due to
the work of Dashen, Manohar and Jenkins [2].  By combining the large $N_c$ 
counting rules and the chiral lagrangian, they pointed out that there is a 
contracted SU(2$N_f$) light quark spin-flavor symmetry in the baryon 
sector in the large $N_c$ limit.  
This symmetry can be also derived in the Hartree-Fock picture
[3, 4].  (Similar result was obtained before [5].)  The observation of the
light quark spin-flavor symmetry results in many quantitative applications
of $1/N_c$ expansion to both light and heavy baryons [2, 6-10].  Within the 
framework of large $N_c$ HQET, we discussed the heavy baryon masses [9],
and emphasized in the Hartree-Fock picture that the heavy baryon mass equals
to the heavy quark mass plus the proton mass in the large $N_c$ limit.  
Actually this point was first pointed out in Ref. [11] by Chow and Wise.
In the Hartree-Fock picture, by observing the light quark dominance in the
large $N_c$ limit, we further deduced that the heavy baryon-pion coupling
constant equals to the light baryon-pion coupling constant, which is a 
result of Jenkins in Ref. [2].  This paper will develop this deduction.  
Our discussion will be not restricted to heavy  baryons.  It applies to both
light and heavy baryons.  And it will not be limited to the baryon 
interaction
with pion only, the interaction can be with $\rho$ meson.  Actually the
discussion considers baryon interactions with any light meson.  
However, it emphasizes the relation between the heavy baryon-meson coupling 
and the light baryon-meson coupling.
\par
\vspace{1.0cm}
   The baryons of the most general interests are in ground state, that means 
the quarks have no orbital angular momentum excitations in the constituent 
picture.  For simplicity, only two flavors of light quarks are 
considered.  In terms of their quantum numbers spin $J$ and isospin $I$, 
they are 
$(I. J) = (\frac{1}{2}, \frac{1}{2}), (\frac{3}{2}, \frac{3}{2}), ...,
(\frac{N_c}{2}, \frac{N_c}{2})$ for light baryons and 
$(I, J) = (0, 0), (1, 1), ..., (\frac{N_c-1}{2}, \frac{N_c-1}{2})$ for heavy
baryons.  In the large $N_c$ limit, because of the light quark spin-flavor 
symmetry, baryons belonging to the same tower of $(I, J)$ are degenerate.
Each tower is an irreducible representation of this symmetry.  As in Ref. 
[9], we have been working in the Hartree-Fock picture for baryons.  Many 
interesting results can be obtained in a simple way.  In the following, 
we just consider the ground state baryons.  However, all the arguments can 
be generalized to the excited baryon cases.  
\par
\vspace{1.0cm}
   The first result is that {\it \bf in the large $N_c$ limit, all the 
coupling 
constants of the baryon interactions with fixed light meson are equal
to each other.}  When we talk about the coupling constant, the possibily 
explicit factors of $N_c$ and the Clebsch-Gordon coefficient have been 
factored out.
Generally, every vertex of the baryon-meson intereactions in a given tower
of $(I, J)$ can be expressed by the fields combination of the 
initial baryon, the final baryon and the light meson
times some coupling coefficient.  The coupling coefficient can be 
parameterized into the product of certain constant and Clebsch-Gordon 
coefficients due to the symmetry group of the interaction.  The constant 
is then called coupling constant (or coupling for short).  
Note that both the initial and the final states of baryons
in the interaction belong to same tower of $(I, J)$.  Consider the light 
baryon case.  The quark spin-flavor symmetry tells us that all the light 
baryon-meson couplings are equal.  And because the interaction is determined 
by the light quarks inside the baryons, the heavy baryon-meson couplings are 
also 
equal to each other due to this symmetry.   In the $N_c\rightarrow \infty$ 
limit, the
heavy baryons are dominated by the light quark systems that also dominate 
the light baryons.  In this case, the heavy baryon-meson coupling constant
equals to the light baryon-meson one.  Therefore all the ground state 
baryon and light meson interaction couplings are the same in the large $N_c$ 
limit.  
Of course, for different light mesons, the couplings should be different.
\par
\vspace{1.0cm}
   Note that it is the observation of the light quark dominance in large
$N_c$ baryons that establishes the large $N_c$ equal relation between the 
heavy baryon-meson coupling and the light baryon-meson coupling in the 
Hartree-Fock picture.  The light quark spin independence [3], or the light
quark spin-flavor symmetry, only gives the coupling equal relation within one
given tower of $(I, J)$.  The "heavy-light" equal relation, however, subjects 
to the $1/N_c$ correction.
\par
\vspace{1.0cm}
   The second result concerns the $1/N_c$ corrections of the above 
conclusion.  $1/N_c$ corrections violate the spin-flavor symmetry.  And we 
note the baryon spectrum has the relation $I=J$.  Therefore the coupling 
of the baryon-meson interaction has the following $1/N_c$ expansion,
\begin{equation}
g=g_0[1+c_1\frac{L^2}{N_c}+c_2\frac{J_1^2+J_2^2}{N_c^2}+
c_3\frac{L^4}{N_c^2}+O(\frac{1}{N_c^3})]~,
\end{equation}
where $g_0$ is the coupling constant in the large $N_c$ limit, which is the
same for both heavy and light baryons.   
$c_i$ $(i=1, 2, 3)$ are unknown coefficients.  
$J_1$ and $J_2$ denote the baryon spins of 
the initial and final states, respectively, and $L$ stands for the light
meson total angular momentum 
in the rest frame of the initial baryon, $\vec{L}=\vec{J}_1-\vec{J}_2$, 
which counts the orbital angular momentum and
the spin of the meson.  When the $1/N_c$ corrections are considered, the 
coupling constant is function of $J_1^2$, $J_2^2$ and $J_1 \cdot J_2$.  The
quantity $J_1 \cdot J_2$, however, can be reexpressed in terms of $J_1^2$, 
$J_2^2$ and $L^2$.  The factor $N_c$ should appear so as to keep the
$N_c$ scaling for $g$.  In the extreme case while in the baryon all the 
quark spins align in the same direction, $J_1^2$ and $J_2^2$ scale as 
$\frac{N_c^2}{4}$.  Only by dividing a factor $N_c^2$, have the terms 
proportional to $J_1^2$ or $J_2^2$ in above equation the right $N_c$ 
scaling.  However, $L^2$ is always a fixed quantity which does not depend
on $N_c$.  So generally this term is suppressed by $N_c$.  
Because strong intereaction is CP invariant, $J_1^2$ and $J_2^2$ terms
have the same coefficient.  
For different towers of $(I, J)$, that means for the light and
heavy baryons, the coefficients $c_i$ are not neccessarily the same.  From
Eq. (1) and above discussion, we see the following points.  (a)  {\it \bf 
Within a given tower of $(I, J)$, the equal relation of the coupling 
constants 
receives a correction actually only at the order of $1/N_c^2$}.  This is 
because within the tower of $(I, J)$, the term $c_1\frac{L^2}{N_c}$ is a
constant.  The couplings can be redefined by absorbing this constant, so 
that there is an explicit equal relation among the redefined couplings, 
which gets the  
corrections from $1/N_c^2$.  Furthermore, the redefinition can be made to 
absorb all the terms of $(L^2)^n$ $(n=1, 2, 3, ...)$, which are constants 
within the tower of $(I, J)$.  In the mixing terms like 
$L^2(J_1^2+J_2^2)/N_c^3$, 
the factor $L^2/N_c$ 
can be absorbed into the coefficients.  Therefore, the coupling constant 
can be expanded only by the powers of $J_1^2/N_c^2$ and $J_2^2/N_c^2$ 
in a given tower of $(I, J)$.  (b)
{\it \bf Between the different towers of $(I, J)$, the equal relation 
receives
correction at the order of $1/N_c$.}  This is simply because the value of 
the 
coefficient $c_1$ in the light baryon case is generally different from 
that in the heavy baryon case.  So their difference gives an order of 
$1/N_c$ correction to the coupling constant equal relation.  
Furthermore in the extreme situation while
$J_1^2$ and $J_2^2$ $\sim N_c^2/4$, in Eq. (1) the subleading terms which 
are proportional to 
$(\frac{J_1^2}{N_c^2})^{m_1}(\frac{J_2^2}{N_c^2})^{m_2}$ ($m_1$ and $m_2$ 
are non-negative integers) have the leading behaviour.  Because
of the light quark dominance, the large $N_c$ equal relation of the 
couplings now gives that the summation of the coefficients of these 
subleading terms for the light baryons equals to that for the heavy 
baryons.  Under the specific assumption that the large $N_c$ limit and the 
heavy quark limit are commutative, we may also have the equal relation of 
the coefficient $c_1$ for the light baryons and the heavy baryons.
\par
\vspace{1.0cm}
   The first result of $1/N_c$ corrections (a) can be obtained directly in 
view of Ref. [4] in the Hartree-Fock picture.  However because we also have had 
the large $N_c$ coupling equal relation between different towers of $(I, J)$,
it is necessary to ask its $1/N_c$ correction.  The discussion on the second
result (b) has made the origin of the $1/N_c$ correction to the "heavy-light"
coupling equal relation clear.  
\par
\vspace{1.0cm}
  For the light meson being a pion, the main conclusions for the 
baryon-meson interactions we have obtained are the same as that originally
obtained in Ref. [2].  However, working in the Hartree-Fock picture, the
way of understanding is simpler, 
and is not subject to the soft pion limit.  
All the results can be immediately
generalized to the baryon interactions with other light meson cases, like
the baryon-$\rho$ meson interactions.  
Similar to baryon-pion intereaction vertex described in Ref. [2], the 
baryon-$\rho$ vertex is
\begin{equation}
\bar{B_2}G^{\rho}_{ai}B_1\rho^{ai},
\end{equation}
where $a$ denotes the isospin of $\rho$ meson and $i$ labels the spin 
component of the $\rho$ meson in the baryon rest frame.  
The coupling coefficient can be written as 
\begin{equation}
<I_2 I_{2z}, J_2 J_{2z}|G^{\rho}_{ai}|I_1 I_{1z}, J_1 J_{1z}>=
N_cg^{\rho}(J_1, J_2)\sqrt{\frac{2J_1+1}{2J_2+1}}
\left(\begin{array}{ccc}
I_1   &1|&I_2\\
I_{1z}&a|&I_{2z}
\end{array}
\right)
\left(\begin{array}{ccc}
J_1   &1|&J_2   \\
J_{1z}&i|&J_{2z}
\end{array}
\right)~,
\end{equation}
where $g^{\rho}(J_1, J_2)$ is the coupling constant which is of order 
$1$ because the $N_c$ dependence has been factored out.  The 
Clebsch-Gordon coefficient is determined from angular-momentum and isospin
conservation.  The coupling constant $g^{\rho}(J_1, J_2)$, which is the
main topic of this Letter, is given by Eq. (1).  In this way, the $NN\rho$ 
and $N\Delta\rho$ intereaction couplings are related to each other,
where $N$ and $\Delta$ denote nucleon and $\Delta$-baryon.  It is 
interesting to note that the $p$-wave $NN\pi$ and $N\Delta\pi$ intereaction
vertice [2] have the same form as that of the $s$-wave $NN\rho$ and
$N\Delta\rho$ intereactions, 
therefore the ratios of the couplings, in which the 
Clebsch-Gordon coefficients are canceled, have the relation
\begin{equation}
\frac{g_{N\Delta\rho}}{g_{N\Delta\pi}}=\frac{g_{NN\rho}}{g_{NN\pi}}~,
\end{equation}
which is valid up to $1/N_c^2$.  Actually this relation has been suggested 
by phenomenological models [12].
Correspondingly, we also have the same relation for heavy baryon case,
\begin{equation}
\frac{g_{\Lambda_Q\Sigma_Q\rho}}{g_{\Lambda_Q\Sigma_Q\pi}}=
\frac{g_{NN\rho}}{g_{NN\pi}}~.
\end{equation}
However, this relation will receive correction at the order of $1/N_c$.
\par
\vspace{1.0cm}
   Finally let us make a comment on the recent experimental result of the 
strong couplings in heavy baryon chiral lagrangians [13, 14].  There are 
two coupling contants, $g_1$ and $g_2$, where the notation of Ref. [13] is 
adopted.  They can be determined from the transitions 
$\Sigma_c^*\rightarrow \Sigma_c\pi$ and $\Sigma_c\rightarrow \Lambda_c\pi$,
respectively.  In the large $N_c$ limit, the equal coupling relation gives
\begin{equation}
|g_1|=\sqrt{2}|g_2|=g_A^N~,
\end{equation}
where $g_A^N\simeq 1.25$ is the light baryon-pion coupling constant.  From 
the CLEO data [15], it was obtained that [16]  
\begin{equation}
|g_2|=0.57\pm 0.10~.
\end{equation}
It deviates from the large $N_c$ expectation remarkably.  But such a deviation
is not surprising, because the relation $|g_2|=\frac{1}{\sqrt{2}}g_A^N~$ will
accept correction at the order of $1/N_c$.  This means that the coefficient 
$c_1$ of Eq. (1) for heavy baryon case is different from that for light baryon
case.  In other words, the large $N_c$ limit and the heavy quark limit are not
commutative.  Another large $N_c$ relation $|g_1|=\sqrt{2}|g_2|$, on the other 
hand, does not receive corrections until to the order of $1/N_c^2$.  Therefore
$|g_1|\simeq 0.81\pm 0.14$ is expected quite accurate.  It will be checked in 
the experiments in the near future.
\par
\vspace{1.0cm}
   In summary, 
the original results for 
the baryon-pion couplings in the large $N_c$ QCD can be understood in 
a simpler way in the Hartree-Fock picture.  The large $N_c$ relation and its 
$1/N_c$ correction between the heavy baryon-meson coupling and the light baryon
meson coupling are stressed.  While the simplification
maybe inspiring, we have made a simple application to the baryon-$\rho$ 
meson intereactions.  We have also discussed implications of the recent 
experimental result for the strong coupling constants of the heavy baryon 
chiral lagrangians.  
\par
\vspace{2.0cm}

   We would like to thank M. Kim and S. Kim for helpful discussions. 
This work was supported by the Korea Science and Engineering Foundation
through the SRC programm.

\end{document}